\documentclass[conference]{IEEEtran}

\usepackage{fixltx2e}
\usepackage{lettrine}
\usepackage{cite}
\usepackage[utf8]{inputenc}
\usepackage{fancyhdr}
\usepackage[detect-weight, product-units = power,inter-unit-product = {}\cdot{}]{siunitx}
\usepackage{tikz}
\usepackage{booktabs}
\usepackage{balance}
\usepackage{graphicx}

\ifCLASSINFOpdf
\else
\fi

\usepackage{amsmath}

\newcommand{\mrx}{\mbox{Mn$_2$Ru$_x$Ga}}
\newcommand{\tcmp}{T_{\text{comp}}}

\begin{document}
\title{Study of the effect of annealing on the\\ properties of \mrx~thin films}
\author{\IEEEauthorblockN{K.E. Siewierska\IEEEauthorrefmark{1},
G. Atcheson\IEEEauthorrefmark{1},
K. Borisov \IEEEauthorrefmark{1},
M. Venkatesan\IEEEauthorrefmark{1},
K. Rode \IEEEauthorrefmark{1} and
J.M.D. Coey\IEEEauthorrefmark{1}}
\IEEEauthorblockA{\IEEEauthorrefmark{1}School of Physics \& CRANN
University of Dublin Trinity College, Dublin 2, Ireland\\ Email: siewierk@tcd.ie, rodek@tcd.ie \& jcoey@tcd.ie} 
}
\maketitle

\begin{abstract}
  The effect of vacuum annealing thin films of the compensated
  ferrimagnetic half-metal \mrx~at
  temperatures from \SIrange{250}{400}{\degreeCelsius} is investigated. The
  \SI{39.3}{\nano\metre} films deposited on (100) MgO substrates exhibit perpendicular
  magnetic anisotropy due to a small $\sim\SI{1}{\percent}$ tetragonal elongation induced
  by substrate strain. The main change on annealing is a modification in the
  compensation temperature $\tcmp$, which first increases from \SI{50}{\kelvin}
  for the as-deposited film to \SI{185}{\kelvin} after annealing at
  \SI{250}{\degreeCelsius}, and
  then falls to \SI{140}{\kelvin} after annealing at \SI{400}{\degreeCelsius}. There are minor
  changes in the atomic order, coercivity, resistivity and anomalous Hall
  effect (AHE), but the net magnetization measured by SQUID magnetometry with
  the field applied in-plane or perpendicular-to-the-plane changes more
  significantly. It saturates at \SIrange{20}{30}{\kilo\ampere\per\metre} at room
  temperature, and a small soft component is seen in the perpendicular SQUID
  loops which is absent in the square AHE hysteresis loops. This is explained
  by the half-metallic nature of the compound; the AHE probes only the $\mathbf{4c}$ Mn
  sublattice that provides the spin-polarized electrons at the Fermi level,
  whereas the SQUID measures the sum of the oppositely-aligned $\mathbf{4c}$ and $\mathbf{4a}$
  sublattice magnetisations.\\
\end{abstract}

\noindent \small \textcopyright 2017 IEEE. Personal use of this material is permitted. Permission from IEEE must be obtained for all other uses, in any current or future media, including
reprinting/republishing this material for advertising or promotional purposes, creating new
collective works, for resale or redistribution to servers or lists, or reuse of any copyrighted component of this work in other works.


\IEEEpeerreviewmaketitle

\section{Introduction}
\lettrine{A}{} dream material for spintronics would have low or zero
spontaneous magnetization $M_s$, produce no stray fields
but be \SI{100}{\percent} spin polarised. This useful combination is
possible in a fully compensated half metallic ferrimagnet, or a
zero-moment half metal (ZHM). Such a material was
envisaged by de Groot in 1983 \cite{deGroot_1983} (he called it a half-metallic
antiferromagnet, although the two sublattices cannot be
crystallographically equivalent), but the first example
\mrx~(MRG) was only discovered experimentally in
2014 \cite{Kurt_MRG}. 

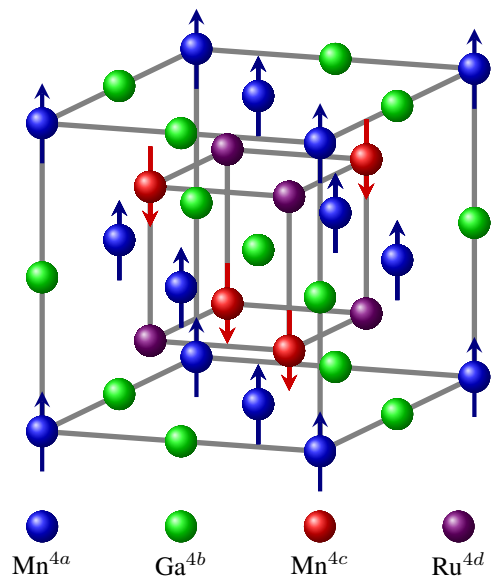
\begin{figure}
    \centering
    \newcommand{\myPgfX}{}%
\newcommand{\myPgfY}{}%
\newcommand{\myPgfZ}{}%
  \begin{tikzpicture}[scale = 4.1, z = -2.50mm, rotate around y = -15, 
    atom/.style = {circle, shade, shading = ball, inner sep = 0, minimum size = 12.0} ]
    \draw [help lines,line width=2pt] (0,0,0) -- (0,0,1) -- (0,1,1) -- (0,1,0) -- (0,0,0) -- (1,0,0) -- (1,0,1) -- (1,1,1) -- (1,1,0) -- (1,0,0) (1,1,0) -- (0,1,0) (0,1,1) -- (1,1,1) (0,0,1) -- (1,0,1);
    \draw [help lines,line width=2pt] (0.25,0.25,0.25) -- (0.25,0.25,0.75) -- (0.75,0.25,0.75) -- (0.75,0.25,0.25) -- cycle -- (0.25,0.75,0.25) -- (0.75,0.75,0.25) -- (0.75,0.75,0.75) -- (0.25,0.75,0.75) -- (0.25,0.75,0.25) (0.25,0.75,0.75) -- (0.25,0.25,0.75) (0.75,0.75,0.75) --(0.75,0.25,0.75) (0.75,0.75,0.25) -- (0.75,0.25,0.25);
    \foreach \A/\B/\C in {0/0/0,1/0/0/,0/1/0,0/0/1,1/1/0,0/1/1,1/1/1,1/0/1}
      \foreach \x/\y/\z in {0/0/0,0/0.5/0.5,0.5/0/0.5,0.5/0.5/0}
        \pgfmathsetmacro{\myPgfX}{{(\A+\x > 1) ? frac(\A+\x) : \A+\x}}%
        \pgfmathsetmacro{\myPgfY}{{(\B+\y > 1) ? frac(\B+\y) : \B+\y}}%
        \pgfmathsetmacro{\myPgfZ}{{(\C+\z > 1) ? frac(\C+\z) : \C+\z}}%
          \draw [->,>=stealth,ultra thick,blue!50!black] (\myPgfX,\myPgfY-0.13,\myPgfZ) -- +(0,0.26,0);
    \foreach \A/\B/\C in {0/0/0,1/0/0/,0/1/0,0/0/1,1/1/0,0/1/1,1/1/1,1/0/1}
      \foreach \x/\y/\z in {0/0/0,0/0.5/0.5,0.5/0/0.5,0.5/0.5/0}
        \pgfmathsetmacro{\myPgfX}{{(\A+\x+0.25 > 1) ? frac(\A+\x+0.25) : \A+\x+0.25}}%
        \pgfmathsetmacro{\myPgfY}{{(\B+\y+0.25 > 1) ? frac(\B+\y+0.25) : \B+\y+0.25}}%
        \pgfmathsetmacro{\myPgfZ}{{(\C+\z+0.25 > 1) ? frac(\C+\z+0.25) : \C+\z+0.25}}%
	  \draw [<-,>=stealth,ultra thick,red!80!black] (\myPgfX,\myPgfY-0.13,\myPgfZ) -- +(0,0.26,0);
    \foreach \A/\B/\C in {0/0/0,1/0/0/,0/1/0,0/0/1,1/1/0,0/1/1,1/1/1,1/0/1}
      \foreach \x/\y/\z in {0/0/0,0/0.5/0.5,0.5/0/0.5,0.5/0.5/0}
        \pgfmathsetmacro{\myPgfX}{{(\A+\x > 1) ? frac(\A+\x) : \A+\x}}%
        \pgfmathsetmacro{\myPgfY}{{(\B+\y > 1) ? frac(\B+\y) : \B+\y}}%
        \pgfmathsetmacro{\myPgfZ}{{(\C+\z > 1) ? frac(\C+\z) : \C+\z}}%
	\node at (\myPgfX,\myPgfY,\myPgfZ) [atom, ball color = blue] {};
    \foreach \A/\B/\C in {0/0/0,1/0/0/,0/1/0,0/0/1,1/1/0,0/1/1,1/1/1,1/0/1}
      \foreach \x/\y/\z in {0/0/0,0/0.5/0.5,0.5/0/0.5,0.5/0.5/0}
        \pgfmathsetmacro{\myPgfX}{{(\A+\x+0.5 > 1) ? frac(\A+\x+0.5) : \A+\x+0.5}}%
        \pgfmathsetmacro{\myPgfY}{{(\B+\y+0.5 > 1) ? frac(\B+\y+0.5) : \B+\y+0.5}}%
        \pgfmathsetmacro{\myPgfZ}{{(\C+\z+0.5 > 1) ? frac(\C+\z+0.5) : \C+\z+0.5}}%
          \node at (\myPgfX,\myPgfY,\myPgfZ) [atom, ball color = green] {};
    \foreach \A/\B/\C in {0/0/0,1/0/0/,0/1/0,0/0/1,1/1/0,0/1/1,1/1/1,1/0/1}
      \foreach \x/\y/\z in {0/0/0,0/0.5/0.5,0.5/0/0.5,0.5/0.5/0}
        \pgfmathsetmacro{\myPgfX}{{(\A+\x+0.25 > 1) ? frac(\A+\x+0.25) : \A+\x+0.25}}%
        \pgfmathsetmacro{\myPgfY}{{(\B+\y+0.25 > 1) ? frac(\B+\y+0.25) : \B+\y+0.25}}%
        \pgfmathsetmacro{\myPgfZ}{{(\C+\z+0.25 > 1) ? frac(\C+\z+0.25) : \C+\z+0.25}}%
          \node at (\myPgfX,\myPgfY,\myPgfZ) [atom, ball color = red] {};
    \foreach \A/\B/\C in {0/0/0,1/0/0/,0/1/0,0/0/1,1/1/0,0/1/1,1/1/1,1/0/1}
      \foreach \x/\y/\z in {0/0/0,0/0.5/0.5,0.5/0/0.5,0.5/0.5/0}
        \pgfmathsetmacro{\myPgfX}{{(\A+\x+0.75 > 1) ? frac(\A+\x+0.75) : \A+\x+0.75}}%
        \pgfmathsetmacro{\myPgfY}{{(\B+\y+0.75 > 1) ? frac(\B+\y+0.75) : \B+\y+0.75}}%
        \pgfmathsetmacro{\myPgfZ}{{(\C+\z+0.75 > 1) ? frac(\C+\z+0.75) : \C+\z+0.75}}%
          \node at (\myPgfX,\myPgfY,\myPgfZ) [atom, ball color = red!50!blue] {};

	  \node at (-0.65+0.15,-0.55) [rotate around y = 15, atom, ball color = blue, label = {below:Mn$^{4a}$}] {};
	  \node at (-0.20+0.15,-0.55) [rotate around y = 15, atom, ball color = green, label = {below:Ga$^{4b}$}] {};
	  \node at (0.25+0.15,-0.55)  [rotate around y = 15, atom, ball color = red, label = {below:Mn$^{4c}$}] {};
	  \node at (0.70+0.15,-0.55)  [rotate around y = 15, atom, ball color = red!50!blue, label = {below:Ru$^{4d}$}] {};
  \end{tikzpicture}
    \caption{Model of the L2$_1$ structure of \mrx.}
    \label{figure:model}
\end{figure}

In the solid solution between the ferrimagnetic end
members Mn$_2$Ga and Mn$_2$RuGa, the extrapolated MRG net moment
is found to change sign at $T = 0$ at $x=0.5$, which corresponds to 21
valence electrons in the cubic L2$_1$ Heusler structure \cite{Kurt_MRG,Graf,Felser_book,Zic}, as shown in \figurename~\ref{figure:model}. This demonstration opened a new field of spin electronics with no
net moment, with potential applications in high frequency oscillators and spin-torque devices \cite{Mitani,Nivetha,Awari,AIP,Betto_MRG,Kiril}. If MRG is to be integrated into a device manufactured
using silicon-based CMOS technology, the MRG layer will
have to be able to survive the relevant processing
conditions without modification of properties, and must be
smooth. A major concern is the potential diffusion of
elements in and from the MRG layer \cite{Hf_paper}. It is also important to
understand how the properties of MRG may change due to
the annealing at various temperatures. More fundamentally,
there is an opportunity to examine the relation between the
net magnetization $M$ and Hall resistivity $\rho_{xy}$ for one
sublattice in a half-metallic ferrimagnet \cite{Kurt_MRG}.
Here we investigate the effect of magnetic annealing on the
properties of the thin films of Mn$_2$Ru$_{0.7}$Ga, in the as-deposited
state, and annealed in vacuum at temperatures up to \SI{400}{\degreeCelsius}.
Under post-deposition annealing the MRG is able to maintain
the high-quality, as-deposited surface roughness.

\section{Methodology}

\subsection{Sample preparation}
An epitaxial thin film of MRG was grown by DC magnetron
sputtering, using our Shamrock sputtering system, on a \SI{25 x
25}{\milli\metre} (100) MgO substrate. The film was co-sputtered in argon
from two \SI{75}{\milli\metre} targets of Mn$_2$Ga and Ru onto the substrate
maintained at \SI{380}{\degreeCelsius}. The chosen composition was
Mn$_2$Ru$_{0.7}$Ga, which exhibits compensation below room
temperature \cite{Kurt_MRG,Graf,Felser_book}. The film was capped \emph{in
situ} with a \SI{3}{\nano\metre}
layer of AlO$_x$ deposited at room temperature in order to
prevent oxidation. The large sample was diced into four equal
squares, and smaller pieces for subsequent treatment and
measurements.

Pieces of the thin film were annealed in a vacuum of \SI{e-6}{\milli\bar} in a
perpendicular magnetic field of \SI{800}{\milli\tesla} for one hour.
The annealing temperatures chosen were \SIlist{250;300;350;400}{\degreeCelsius}. Provided the annealing temperature is kept below
\SI{400}{\degreeCelsius}, the L2$_1$ crystal structure of MRG is maintained and Mn
diffusion out of the film is minimized \cite{Gavin}.

\subsection{Structural properties}

A Bruker D8 X-ray diffractometer with a copper
tube emitting K$_{\alpha 1}$ X-rays with wavelength \SI{154.06}{\pico\metre} and a double-bounce Ge [220] monochromator was used
to determine the diffraction patterns of the thin films. Low angle
X-ray reflectivity was measured using a Panalytical
X’Pert Pro diffractometer, and thickness was found by fitting
the interference pattern using X’Pert Reflectivity software with a least square fit.

\subsection{Electrical properties}

Longitudinal and Hall Resistivity were measured using the
4-point Van der Pauw method with indium contacts and an
applied current of \SI{5}{\milli\ampere}.

\subsection{Magnetic properties}

Measurements of the magnetization with magnetic field
applied perpendicular or parallel to the surface of the films
was carried out using a \SI{5}{\tesla} Quantum Design SQUID.
Hysteresis loops were measured in fields up to \SI{5}{\tesla}, at
temperatures between \SI{100}{\kelvin} and room temperature. These
data were corrected for the diamagnetism of the substrate.
Thermal scans in \SI{30}{\milli\tesla} after saturation of the magnetization
at room temperature were used to determine the compensation
temperatures of the annealed and unannealed films. 

\section{Results}

\subsection{Structural Properties: Crystal Structure and Thickness}

\begin{figure}
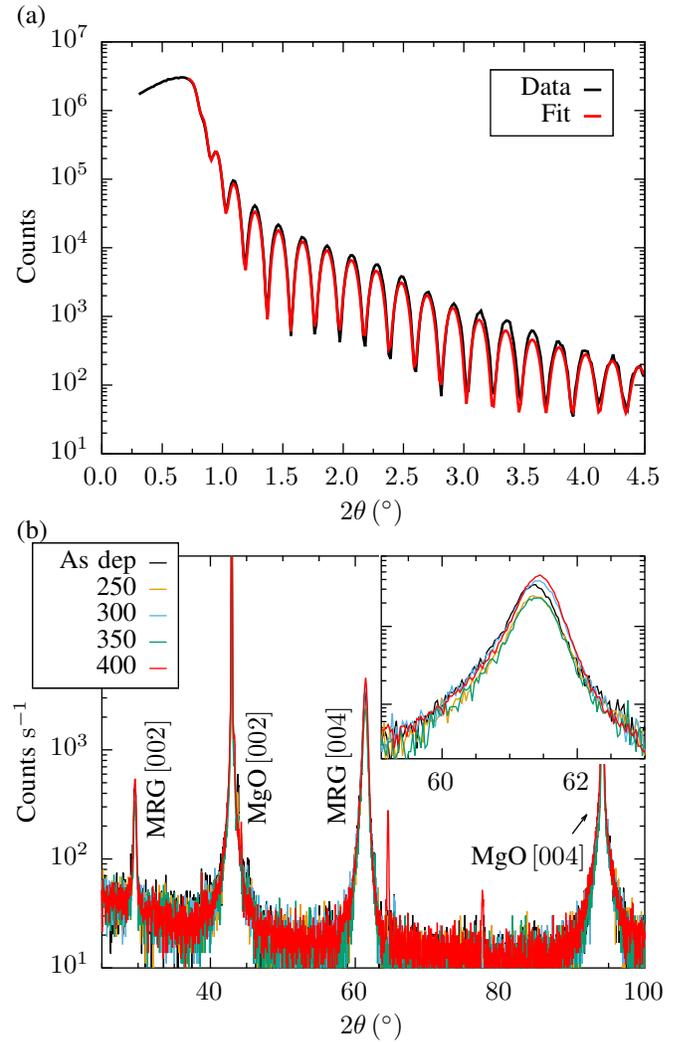

    \centering
    \input{./fig2a.tex}
    \input{./fig2b.tex}
    \caption{(a) XRR of the as deposited Mn$_2$Ru$_{0.7}$Ga thin film. Fitting for a thickness of \SI{39.3}{\nano\metre} and a density of \SI{8.3}{\gram\per\centi\metre\cubed} (b) XRD pattern of Mn$_2$Ru$_{0.7}$Ga thin films on MgO substrate at different anneal temperatures. Insert shows the (004) peak of MRG.}
    \label{figure:xrays}
\end{figure}

X-ray data on the films are shown in \figurename~\ref{figure:xrays}. Fitting the X-ray
reflectivity in \figurename~\ref{figure:xrays}(a) gives a film thickness of
\SI{39.3}{\nano\metre}. There
are no significant differences among four samples taken from
different parts of the large film. The diffraction patterns in \figurename~\ref{figure:xrays}(b)
of the as-deposited Mn$_2$Ru$_{0.7}$Ga film and samples annealed
at \SIlist{250;300;350;400}{\degreeCelsius} in the perpendicular
magnetic field exhibit (002) and (004) reflections from the
MRG, together with peaks from the MgO substrate. There are
only small changes in the relative (002) and (004) peak intensities
($I_{(002)}/I_{(004)}$) on annealing, in the range \numrange{0.11}{0.14}
(\tablename~\ref{table_data}). This small value is indicative of a high degree
of atomic order in the L2$_1$ Heusler structure [3]. The broadening of the (002) reflection is
consistent with the measured MRG film thickness. Peak shifts
on annealing are very small, and the $c$ parameter of \SI{604.2}{\pico\metre}
decreases by only about \SI{0.6}{\pico\metre} (\tablename~\ref{table_data}).

\begin{figure}
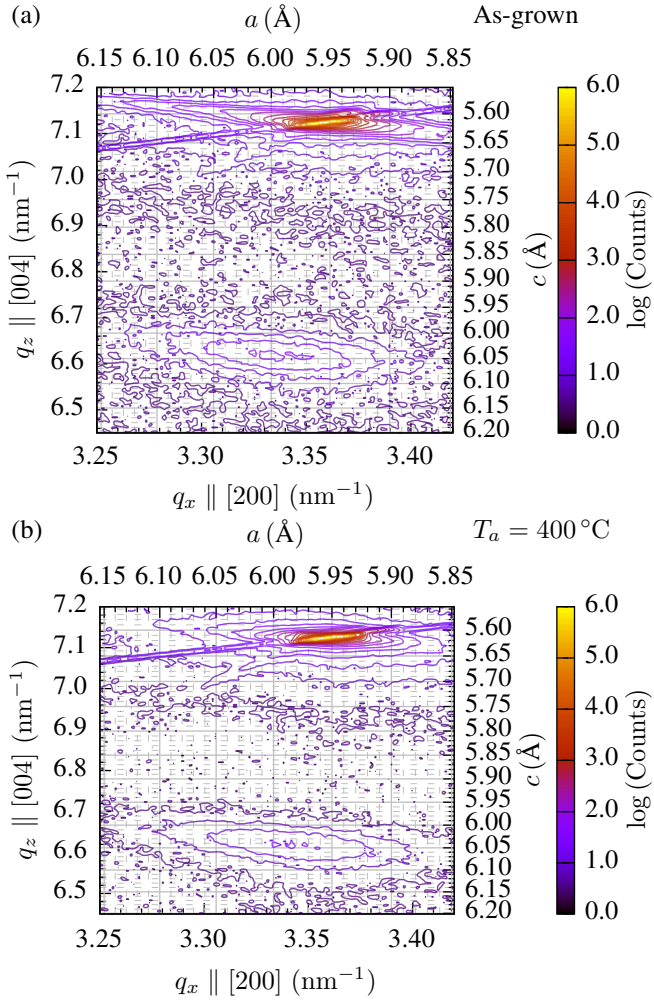

    \centering
    \input{./rsmRT.tex}
    \input{./rsm400.tex}
    \caption{RSM of MgO (113) peak and (a) as-deposited MRG (204) peak and (b) MRG annealed at T$_a={\SI{400}{\degreeCelsius}}$ (204) peak. The lattice parameters are calculated with respect to the MRG unit cell.}
    \label{figure:RSM}
\end{figure}

A reciprocal space map of the as-deposited film, \figurename~\ref{figure:RSM}(a),
confirms the $c$ parameter, and shows a distribution of a
parameters around the central value of \SI{595.8}{\pico\metre}, which corresponds to that
of MgO, to \SI{604}{\pico\metre}. The map for the film annealed at
\SI{400}{\degreeCelsius} is
very similar (\figurename~\ref{figure:RSM}(b)), showing that the crystal structure is
practically unchanged by annealing. The $\sim\SI{1}{\percent}$ substrate-induced
tetragonal expansion of the cubic L2$_1$ cell is
responsible for the perpendicular magnetic anisotropy of the
films.
\subsection{Electrical Properties: Resistivity and Hall Angle}

\begin{figure}
    \centering
    \input{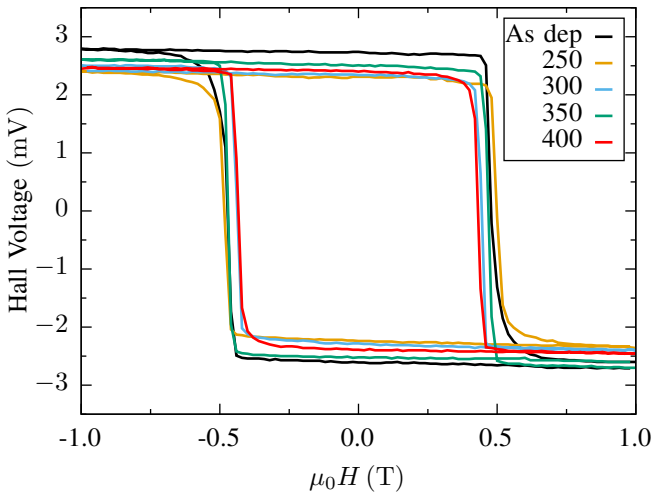}
    \caption{Anomalous Hall effect in MRG annealed at various temperatures. The data have been vertically centered.}
    \label{figure:AHE}
\end{figure}

For Mn$_2$Ru$_{0.7}$Ga films, the calculated
longitudinal $\rho_{xx}$ and
anomalous Hall $\rho_{xy}$ resistivity values show no significant
variation on annealing. The Hall angle was \SI{3.4}{\percent}, and did not
vary significantly from sample to sample. The almost
perfectly-square anomalous Hall loops are shown in \figurename~\ref{figure:AHE},
where the coercivity is close to \SI{450}{\milli\tesla}, regardless of
annealing temperature.

\subsection{Magnetic properties}

\begin{figure}
    \centering
    \input{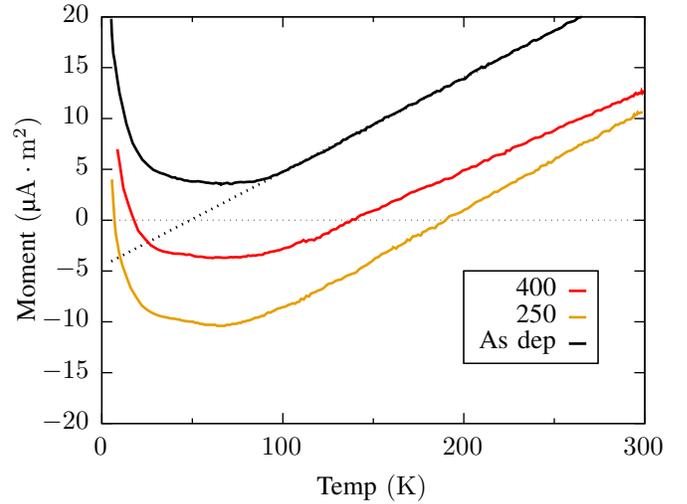}
    \caption{Temperature dependence of moment, measured at remanence (small applied field of \SI{30}{\milli\tesla}) of as-deposited MRG and MRG annealed at \SI{250}{\degreeCelsius} and \SI{400}{\degreeCelsius}.}
    \label{figure:remanence}
\end{figure}

The compensation temperature of MRG is determined from
thermal scans of the remanence after saturating in \SI{5}{\tesla} at room
temperature and scanning in a small field of \SI{30}{\milli\tesla}, performed on three of the samples which are as-deposited MRG and MRG annealed at \SI{250}{\degreeCelsius} and \SI{400}{\degreeCelsius}. Data in
\figurename~\ref{figure:remanence} show a compensation temperature $\tcmp$ of about
\SI{50}{\kelvin} for the as-deposited, unannealed sample. There is a large
Curie-law upturn coming from a few ppm of Fe$^{2+}$ in the MgO
substrate and the magnetization of the film therefore does not
cross zero. Compensation shifts to \SI{185}{\kelvin} for a sample
annealed at \SI{250}{\degreeCelsius}, and is at \SI{140}{\kelvin} in the sample annealed at
\SI{400}{\degreeCelsius} (\tablename~\ref{table_data}). Since the measurements are
performed in a small magnetic field, it is expected that the small diamagnetic
contribution may lead to an overestimation of $\tcmp$. The magnetic moment of the substrate was found to be about \SI{-24.5}{\pico\ampere\metre\squared} and the shift in $\tcmp$ due to this moment is about \SI{0.15}{\kelvin}, which is negligible.  

\begin{figure*}
    \centering
    \input{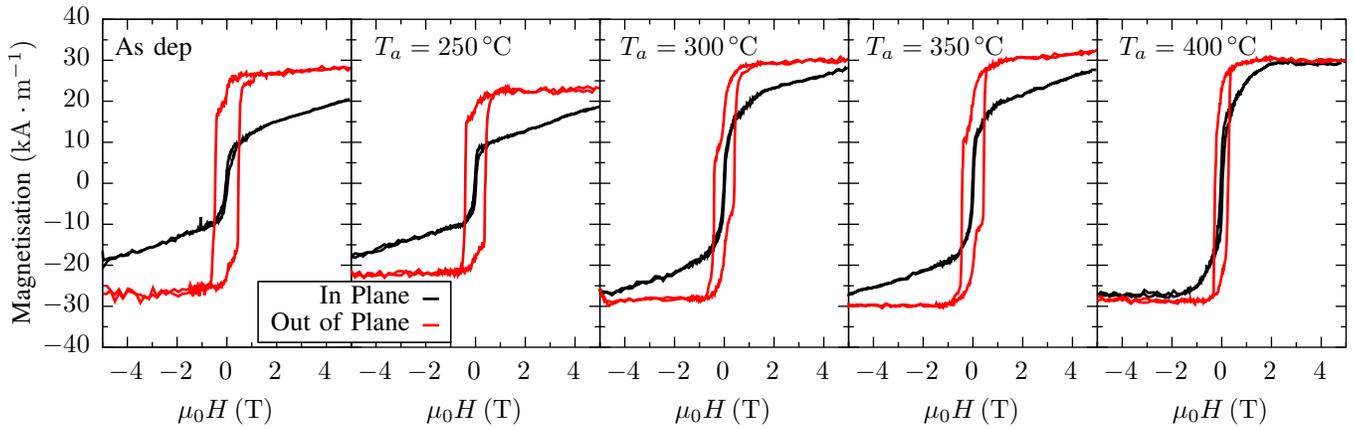}
    \caption{Magnetisation versus applied field, perpendicular and parallel to the surface of the thin films.}
    \label{figure:SQUID}
\end{figure*}

In the room-temperature SQUID measurements (\figurename~\ref{figure:SQUID}),
the as-deposited sample exhibits a square perpendicular
hysteresis loop with coercivity of \SI{466}{\milli\tesla}, but with clear signs
of an easily-saturating component which was not seen in the
AHE data of \figurename~\ref{figure:AHE}. The parallel magnetization also shows a
small, easily-saturated in-plane component and the net
magnetization approaches saturation in fields of order \SI{5}{\tesla}.
The magnetocrystalline anisotropy constant $K_u$ is given by
$K_u = {}^1\!/_2 \mu_0 H_a M_s$
where $\mu_0$ is the permeability in vacuum, $H_a$ is the anisotropy field and
$M_s$ is saturation magnetisation. Assuming an anisotropy field of
\SI{5}{\tesla} and a net magnetization of \SI{20}{\kilo\ampere\per\metre}, we find $K_u$ to be \SI{50}{\kilo\joule\per\metre\cubed}. After annealing
at \SI{250}{\degreeCelsius}, the coercivity decreases
and the hysteresis is squarer, although the shape of the loop
changes with temperature. The easily-saturated component is larger and
appears in the perpendicular loops of the films annealed at
\SIlist{300;350}{\degreeCelsius}, but vanishes in the film annealed at
\SI{400}{\degreeCelsius}. There the coercivity is only \SI{265}{\milli\tesla}, or just over half
of the original value, and strikingly different from that
measured by AHE in smaller fields (\tablename~\ref{table_data}). The easily saturated
in-plane component increases in magnitude with
annealing, becoming almost equal to the out-of-plane moment.

The overall decrease in coercivity and the features of the hysteresis loops may
be related to the variation in concentration of Ru across the thickness MRG
film, with more Ru found closer to the capping layer as found by transmission
electron microscopy \cite{Zic}. This may contribute to
the spread of $a$ in the reciprocal space maps.

\begin{table}
  \begin{minipage}{\columnwidth}
\renewcommand{\arraystretch}{1.3}
\caption{Data for as-deposited and annealed Mn$_2$Ru$_{0.7}$Ga films}
\label{table_data}
\centering
\begin{tabular}{c c c c c c c }
  \toprule
$T$\textsubscript{a} &
$\frac{I_{(002)}}{I_{(004)}}$ &
$c$ & 
$H_{c}^{M(H)}$ &
$H_{c}^{\mathrm{AHE}}$ &
$\frac{H_{c}^{\mathrm{AHE}}}{H_{c}^{M(H)}}$ &
$\tcmp$ \\
(\si{\degreeCelsius}) &
&
(\si{\pico\metre}) &
(\si{\milli\tesla}) &
(\si{\milli\tesla}) &
&
(\si{\kelvin}) \\
\midrule
- & 0.140 & 604.1 & 466 & 485 & 1.04 & 50\\
250 & 0.143 & 604.4 & 413 & 495 & 1.20 & 185\\
300 & 0.119 & 603.6 & 406 & 441 & 1.09 & -\\
350 & 0.145 & 603.8 & 432 & 475 &  1.10 & -\\
400 & 0.119 & 603.6 & 265 &  434, 207\footnote{After saturation in
  \SI{5}{\tesla}} & 1.64 & 140\\
\bottomrule
\end{tabular}
\end{minipage}
\end{table}

\section{Discussion}
Based on the X-ray results in \figurename~\ref{figure:xrays}(b), there seem to be only
minor changes in the atomic order on annealing. However, it
should be emphasized that the substrate temperature during
deposition is \SI{380}{\degreeCelsius}, and the post-deposition annealing for an
hour is mostly carried out at a lower temperature. The
magnetic parameter most sensitive to annealing is $\tcmp$. It is around \SI{50}{\kelvin} in the as-deposited film, rising to
\SI{185}{\kelvin} after annealing at \SI{250}{\degreeCelsius}, and then falling back to
\SI{140}{\kelvin}. Since we are unlikely to significantly alter the Mn($4a$) – Mn($4c$)
exchange coupling by annealing as the lattice parameters
remain unchanged, these large variations in $\tcmp$ must reflect
small changes in manganese populations of the two sites.
Initial annealing might be expected to modify the number of
Ga/Mn($4a$) antisites, thereby increasing the $4c$ sublattice
moment and increasing the compensation temperature \cite{Zic}.
Further annealing at higher temperature (\SI{400}{\degreeCelsius}) produces a
small decrease in $c$, reducing $\tcmp$ again.

Perhaps the most interesting feature of the magnetic data is
the difference between the hysteresis measured by AHE (\figurename~\ref{figure:AHE}) and by
SQUID (\figurename~\ref{figure:SQUID}). Of course, in the half-metallic
ferrimagnetic Heusler compound, one sublattice, $4c$,
contributes predominantly to the AHE signal \cite{Kurt_MRG,Nivetha,Kiril}. The SQUID, on the other hand, measures a small difference of sublattice moments. The Mn moments are
approximately \SI{3}{\mathrm{\mu_B}} per atom, yet the room-temperature
magnetization of order \SI{30}{\kilo\ampere\per\metre} correspond to a net moment
of only \SI{0.05}{\mathrm{\mu_B}} per formula unit. The easily-saturated
component seen there has a moment of only a few percent of a
Mn sublattice moment. This signal does not behave as
expected for traces of ferromagnetic impurity that escaped
detection in XRD. It is linked to the MRG sublattice structure.
A similar component was observed earlier in the
magnetization of D0\textsubscript{22} Mn$_3$Ga \cite{KRode_mn3ga} and Mn$_3$Ge \cite{Kurt_Mn3Ge}, and it is
associated with an in-plane component of the $4a$ ($2b$ in the
D0\textsubscript{22} structure) site moment detected by neutron diffraction
\cite{KRode_mn3ga}. The $4a$ ($2b$) site appears to have easy-cone anisotropy.

The difference in coercivity of the net magnetization (\figurename~\ref{figure:SQUID}) and the
$4c$ sublattice magnetization (\figurename~\ref{figure:AHE}) is striking,
especially in the sample annealed at \SI{400}{\degreeCelsius}. It must be pointed
out that we have used different saturation fields in the two
meaurements, and if we measure the AHE loop in \SI{5}{\tesla}, we find
a coercivity of \SI{207}{\milli\tesla}, which is closer to the \SI{5}{\tesla} SQUID
result. It is unusual to find a minor loop coercivity, which
exceeds that of the major loop (\tablename~\ref{table_data}). We must look for an
explanation that is consistent with the near-compensated
ferrimagnetic structure.
\balance 
\section{Conclusion}
Our studies of the effects of annealing the compensated
ferrimagnetic half metal have shown that the changes in lattice
parameter are $< \SI{0.2}{\percent}$, with minor modifications of the
perpendicular anisotropy. The compensation temperature can
vary however by as much as \SI{130}{\kelvin}. Since $\tcmp$~depends on
the balance between the two manganese sublattices, it is very
sensitive to slight changes in site population due to
redistribution of magnetic Mn and nonmagnetic Ga or Ru
atoms on the two different sites. The issue of the diffusion is addressed elsewhere \cite{Hf_paper}.

The $4c$ sublattice moment shows near-perfect square minor
loops in the \SI{1}{\tesla} AHE measurements, but this does not persist
after saturation in \SI{5}{\tesla}. The hysteresis of the net magnetization
measured by SQUID, which reflects the difference of the two
sublattice moments, is more complex and it reflects the
misalignment of the $4a$ moment with the $c$ axis.

The minor changes in structure of MRG and its AHE
measured in small fields bodes well for its use in devices
annealed at temperatures up to \SI{350}{\degreeCelsius}.

\section*{Acknowledgment}
KE Siewierska \& JMD Coey acknowledge funding through Irish
Research Council and Science Foundation Ireland, grant number 12/RC/2278.
K Rode acknowledges funding through the TRANSPIRE project.

\bibliographystyle{IEEEtran}
\bibliography{refs_intermag}

\end{document}